# Superconducting Ternary Hydridies in Ca-U-H under High Pressure

Juefei Wu[1], Bangshuai Zhu[1], Chi Ding[3], Cuiying Pei[1], Qi Wang[1,2], Jian Sun[3*], Yanpeng Qi[1,2,4*]

1. School of Physical Science and Technology, ShanghaiTech University, Shanghai 201210, China
2. ShanghaiTech Laboratory for Topological Physics, ShanghaiTech University, Shanghai 201210, China
3. National Laboratory of Solid State Microstructures, School of Physics and Collaborative Innovation Center of Advanced Microstructures, Nanjing University, Nanjing 210093, China
4. Shanghai Key Laboratory of High-resolution Electron Microscopy, ShanghaiTech University, Shanghai 201210, China

* Correspondence should be addressed to Y.P.Q. (qiyp@shanghaitech.edu.cn) or J.S. (jiansun@nju.edu.cn)

## Abstract

The research on hydrogen-rich ternary compounds attract tremendous attention for it paves new route to room-temperature superconductivity at lower pressures. Here, we study the crystal structures, electronic structures, and superconducting properties of the ternary Ca-U-H system, combining crystal structure predictions with *ab-initio* calculations under high pressure. We found four dynamically stable structures with hydrogen clathrate cages: $CaUH_{12}$-*Cmmm*, $CaUH_{12}$-*Fd*-$3m$, $Ca_2UH_{18}$-*P*-$3m1$, and $CaU_3H_{32}$-*Pm*-$3m$. Among them, the $Ca_2UH_{18}$-*P*-$3m1$ and $CaU_3H_{32}$-*Pm*-$3m$ are likely to be synthesized below 1 megabar. The *f* electrons in U atoms make dominant contribution to the electronic density of states around the Fermi energy. The electron-phonon interaction calculations reveal that phonon softening in the mid-frequency region can enhance the electron-phonon coupling significantly. The $T_c$ value of $Ca_2UH_{18}$-*P*-$3m1$ is estimated to be 57.5-65.8 K at 100 GPa. Our studies demonstrate that introducing actinides into alkaline-earth metal hydrides provides possibility in designing novel superconducting ternary hydrides.

## Introduction

The pursuit for room temperature superconductivity has been the ultimate goal in the field of superconducting research. The metallization of hydrogen is considered as the potential route, since previous reports predict its superconducting transition temperature $T_c$ to be in the range of 100-760 K under extreme pressures [1, 2]. However, the ultrahigh pressure confines the realization of metal hydrogen in experiments [3-5]. In 2004, Ascroft proposed that comparable high temperature superconductivity could be achieved in hydrogen rich compounds by "chemical precompression" of other elements, which produce valence density for hydrogen metallization at lower pressures [6]. Inspired by this principle, many theoretical studies predict a serious of hydrogen-rich compounds, and some of them are successfully synthesized in the laboratory [7-9]. In particular, $H_3S$ has been found superconducting below 203 K under 155 GPa [10, 11] after theoretical predictions [12, 13], in which the H atoms are covalent bonded with S atoms. Following this milestone, more binary hydrides with clathrate structures have been predicted to have higher $T_c$ values, such as $CaH_6$, $YH_9$, $LaH_{10}$, etc. [14-21], and the $T_c$ value of $LaH_{10}$ is of 250-260 K at 180 GPa, which is close to room temperature [22]. These clathrate structures are highly symmetric, and the H atoms are electron acceptor, forming cages around the guest metal atoms. The geometry is a crucial factor with respect to bonding, electron phonon coupling (EPC) and superconducting properties [23]. Although the stable pressure (>150 GPa) is lower than that for metallization of pure hydrogen, the synthesizing remains challenging. Thus, the next step for superconducting hydrides studies is to further lower the the pressure as well as keeping the high $T_c$ [24].

The search in ternary hydrides is promising among various routes to reduce the stable pressure. The elements combination can be two orders more in ternary hydrides than binary hydrides [23]. This enhances the complexity in both theoretical predictions and experiment synthesis, but also brings about favorable properties. For example, Sun *et al*. predicted a Li doping molecular Mg-H phase $Li_2MgH_{16}$ to reach $T_c$ value 473 K at 250 GPa [25]; Gao *et al*. intercalated a fcc lattice of K with $BH_4$ tetrahedron and the $T_c$ value of $KB_2H_8$ is 134-146 K at 12 GPa [26]; Liang *et al*. predicted the $T_c$ value of $LaBH_8$ to be 156 K at 55 GPa [27]. Moreover, Zhang *et al*. identified a fluorite-type backbone in compositions of the form $AXH_8$ and predicted the $Fm$-$3m$-$LaBeH_8$ to have $T_c$ value of 185 K at 20 GPa [24]; Liu *et al*. summarized generic rules in clathrate structures and proposed representative examples of $CaHfH_{12}$ ($T_c \sim 360$ K at 300GPa) and $CaZrH_{12}$ ($T_c \sim 290$ K at 200GPa) [28]. Besides, Semenok *et al*. synthesized La-Y-H and the $T_c$ matches well with the predictions under high pressure [29]. Therefore, the moderate choice of the elements and appropriate hydrogen clathrate structure designing strategies are two key factors for searching novel superconducting ternary hydrides.

$CaH_6$ is stable at 150 GPa with high $T_c$ value of 220-235 K [14, 30]. Recently, Liang *et al*. predicted $CaYH_{12}$ based on the clathrate structure of $CaH_6$ [31], which later encouraged Zhao *et al*. to find more ternary hydrides in Ca-Y-H system [32]. In addition, the *f* electrons in actinide hydrides are capable of stabilizing hydrogen rich structures at low pressure, such as the confirmation of $UH_{8+\delta}$ in high pressure experiments under 42 GPa [33], and the clathrate structure of $UH_8$ helps the structure designing in $AXH_8$ [24]. Based on the above information, we propose to introduce the actinides element uranium into the alkaline-earth metal hydrides Ca-H under high pressure. This may predict superconducting ternary hydrides with clathrate structures and reduce the stable pressure.

In this work, we focus on exploring the potential superconducting ternary $Ca_xU_yH_z$ hydrides based on the ratio (x+y) : z = 1:6 and 1:8 within the pressure range from 100 GPa to 200 GPa, combining the machine learning graph theory accelerated crystal structure search and first-principles calculations. We found three compositions with dynamically stable structures: $CaUH_{12}$, $Ca_2UH_{18}$ and $CaU_3H_{32}$. $Ca_2UH_{18}$-*P*-3*m*1 and $CaU_3H_{32}$-*Pm*-3*m* are meta-stable and have potential to be synthesized below 1 megabar, while $CaUH_{12}$-*Cmmm* and $CaUH_{12}$-*Fd*-3*m* may exist in annealing process after laser heating. Then, we calculated the electronic properties of the predicted structures. The density of states (DOS), crystal orbital Hamilton population (COHP) and electron localization function (ELF) reveals the contribution of the *f* electrons in U atoms around the Fermi energy. Finally, we calculated the superconducting properties of the predicted structures. The $T_c$ value of $Ca_2UH_{18}$-*P*-3*m*1 is estimated to be 57.5-65.8 K, which approaches the liquid nitrogen level.

Method

We used the machine learning graph theory accelerated crystal structure search method (Magus) [34, 35] to explore the ternary hydrides with thermodynamic stability in $Ca_xU_yH_z$ [ (x+y) : z=1 : 6 or 1 : 8, 1 ≤ x+y ≤ 4 ] at 100 GPa, 150 GPa and 200 GPa. The cutoff energy of the plane-wave was set to 400 eV and the sampling grid spacing of the Brillouin zone was $2\pi \times 0.05$ Å$^{-1}$ in structure searching. To establish the ternary phase diagram, we recalculated the candidate structures and the previously reported structures using the Vienna *Ab-initio* Simulation Package (VASP) based on the density functional theory [36, 37]. The exchange-correlation functional was treated by the generalized gradient approximation of Perdew, Burkey, and Ernzerhof [38]. The calculations used projector-augmented wave (PAW) approach [39] to describe the core electrons and their effects on valence orbitals. We set the plane-wave kinetic-energy cutoff to 800 eV, and the Brillouin zone was sampled by the Monkhorst-Pack scheme of $2\pi \times 0.03$ Å$^{-1}$. The convergence tolerance was $10^{-6}$ eV for total energy and 0.003 eV/Å for all forces. The electronic structure calculations used a denser *k*-mesh grid of $2\pi \times 0.02$ Å$^{-1}$ and the total energy was converged to be less than $10^{-8}$ eV. To study the chemical bonding properties, we performed the COHP analysis using the LOBSTER 4.1 package [40] with the pbeVaspFit2015 basis set.

The phonon spectrum and electron-phonon coupling (EPC) coefficients were calculated by the QUANTUM-ESPRESSO (QE) package [41] using density-functional perturbation theory [42]. We selected the ultrasoft pseudopotential with a kinetic energy cutoff of 100 Ry. The *k*-meshes were $2\pi \times 0.02$ Å$^{-1}$ for the electronic Brillouin zone integration and *q*-meshes of $2\pi \times 0.08$ Å$^{-1}$ for the phonon calculations. The self-consistent solution of the Eliashberg equation [43] and Allen-Dynes modified McMillan equation [44] were used to estimate the superconducting transition temperature $T_c$. The phonon spectrum results were rechecked by the PHONOPY [45] program package using the finite displacement method. The supercell is $2 \times 2 \times 2$ for all the candidate structures and the results agreed with QE calculations.

Results and Discussion

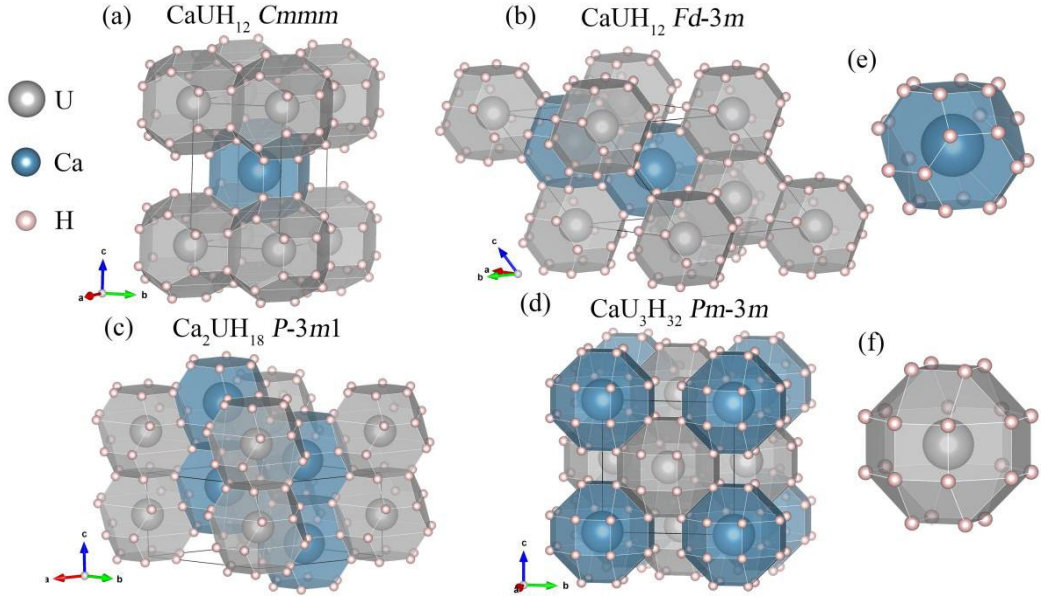

FIG. 1. The predicted crystal structures. (a) $CaUH_{12}$-$Cmmm$ at 80 GPa, (b) $CaUH_{12}$-$Fd$-$3m$ at 150 GPa, (c) $Ca_2UH_{18}$-$P$-$3m1$ at 100 GPa and (d) $CaU_3H_{32}$-$Pm$-$3m$ at 90 GPa. The $H_{24}$ cages in the predicted structures contain (e) quadrangles and hexagons or (f) triangles and quadrangles.

Based on the structure predictions in U-H and Ca-H under high pressure [14, 33], we chose the compositions 1:6 and 1:8 to perform the ternary hydrides structure searches in $Ca_xU_yH_z$ [ (x+y) : z = 1 : 6 or 1 : 8 ]. Each structure evolution was implemented for 30 generations with more than 1,000 structures, we picked out structures with energies not higher than the convex hull by 100 meV/atom, compared their enthalpy under different pressures and calculated their phonon spectrum to examine their dynamical stability. Three compositions $CaUH_{12}$, $Ca_2UH_{18}$ and $CaU_3H_{32}$ have dynamically stable structures, the crystal structures are shown in Fig. 1(a)-(d) and the detailed lattice parameters and atomic coordinates are listed in Table SI. In these four hydrides, hydrogen atoms surround metal atoms forming $H_{24}$ cages. They stack with each other constituting the typical clathrate structures. The $H_{24}$ cages in $CaUH_{12}$-$Cmmm$, $CaUH_{12}$-$Fd$-$3m$ and $Ca_2UH_{18}$-$P$-$3m1$ contain the quadrangles and hexagons [Fig. 1(e)], while the $H_{24}$ cages in $CaU_3H_{32}$-$Pm$-$3m$ compose triangles and quadrangles [Fig. 1(f)]. We compared the average H-H bond lengths of the four predicted structures with $CaH_6$ and $UH_8$ at 120 GPa in Fig. S1. All the lengths are longer than that of the free $H_2$ molecule (0.74 Å). The H-H bond lengths of $CaUH_{12}$-$Cmmm$, $CaUH_{12}$-$Fd$-$3m$ and $Ca_2UH_{18}$-$P$-$3m1$ are close to $CaH_6$, while $CaU_3H_{32}$-$Pm$-$3m$ is close to $UH_8$. This suggests that the structure of ternary hydrides $Ca_xU_yH_z$ with (x+y) : z = 1 : 6 is similar to $CaH_6$ and $Ca_xU_yH_z$ with (x+y) : z = 1 : 8 is analogous to $UH_8$. The phonon spectrum results are in Fig. S2, the absence of imaginary frequencies in the whole Brillouin zone suggest the dynamical stability of the predicted structures within the calculated pressure range. The dynamically stable pressure is 80 GPa for CaUH$_{12}$-*Cmmm*; 150 GPa for CaUH$_{12}$-*Fd-3m*; 100 GPa for Ca$_2$UH$_{18}$-*P-3m*1; 90 GPa for CaU$_3$H$_{32}$-*Pm-3m*. CaUH$_{12}$-*Cmmm* and CaUH$_{12}$-*Fd-3m* have the same composition and their relative enthalpy difference is in Fig. S3. Although the enthalpy of CaUH$_{12}$-*Fd-3m* is above CaUH$_{12}$-*Cmmm*, the enthalpy difference is less than 1 meV/atom above 150 GPa. Such difference is beyond the convergence limitation of our calculations, indicating that *Fd-3m* is possible to be synthesized under high pressure condition.

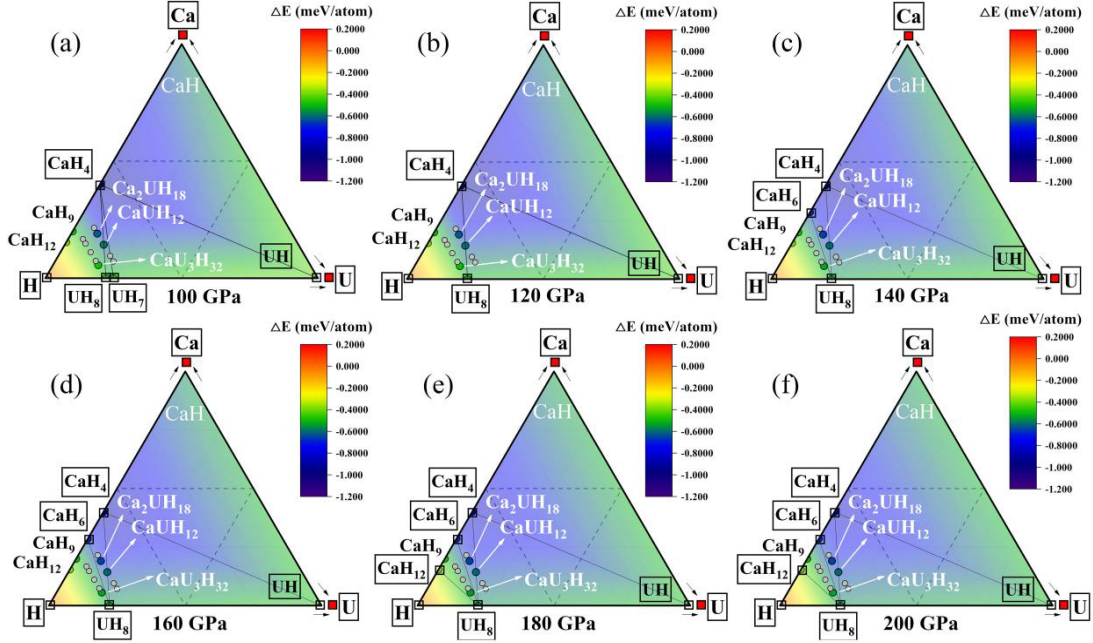

FIG. 2. The ternary convex hull of the Ca-U-H system at different pressures, which starts from the X:H=1:1 (X=Ca, U) site. The compositions in the box are on the ternary convex hull. The solid gray dots indicate the compositions have no stable structures.

To further ensure the thermodynamic stability of the Ca-U-H ternary hydrides, we selected the simple substance Ca, U and H [46-50] under the corresponding pressure as the references as well as the known stable binary hydrides Ca-H [14] and U-H [33] to construct the ternary enthalpy convex hull. To our knowledge, there is no report on the Ca-U binary compounds under high pressure, we performed variable compositions structure searches under 100 GPa, 150 GPa and 200 GPa, respectively. The enthalpies of all the predicted compounds were larger than that of Ca plus U, suggesting that the Ca and U are immiscible above 100 GPa, and no compositions were on the Ca-U side of the ternary convex hull. The Ref. 51 and Ref. 52 pointed out that the zero-point energy (ZPE) has slight effects on the total energy and the overall phase stabilities in Ca-H and U-H under high pressure. Hence, we assume that it is safe to neglect the ZPE contribution in Ca-U-H calculations.

The enthalpy convex hull for Ca-U-H ternary hydrides are shown in Fig. 2. Since the stable compositions are in the hydrogen-rich region, the ternary convex hull starts from the X:H = 1:1 (X =

Ca, U) sites. Despite the predicted structures in $CaUH_{12}$, $Ca_2UH_{18}$ and $CaU_3H_{32}$ are below the simple substance references, they are not on the ternary convex hull [Fig. 2]. Then, we calculated the enthalpy difference between the convex hull and the predicted compositions, as listed in Table I.

Table I. The enthalpy difference relative to the convex hull for $CaUH_{12}$, $Ca_2UH_{18}$ and $CaU_3H_{32}$ under specific pressures.

| | △E (meV/atom) | | | | | |
|---|---|---|---|---|---|---|
| Pressure (GPa) | 100 | 120 | 140 | 160 | 180 | 200 |
| $CaUH_{12}$ | 134.18 | 132.61 | 127.44 | 122.47 | 118.52 | 116.92 |
| $Ca_2UH_{18}$ | 82.33 | 68.90 | 11.60 | 16.70 | 22.40 | 26.78 |
| $CaU_3H_{32}$ | 30.84 | 25.68 | 5.05 | 8.11 | 12.10 | 17.80 |

According to the triangle straight-line method (TSLM), the enthalpy difference of $CaUH_{12}$ is compared with $CaH_4$ and $UH_8$. Although the enthalpy difference of $CaUH_{12}$ lies around 134 meV/atom above the convex hull, it has decreasing trend with the increasing pressure. Despite this value is above the threshold (50 meV/atom) for metastable structures prediction in Ref. 53, some ternary metal compounds can by synthesized within the threshold of 200 meV/atom [54]. Considering the dynamic stability of $CaUH_{12}$-*Cmmm* at 80 GPa and $CaUH_{12}$-*Fd*-3*m* at 150 GPa, they may exist in the quick annealing process after laser heating under high pressure [55]. The maximum enthalpy difference for $CaU_3H_{32}$-*Pm*-3*m* compared to the convex hull is around 31 meV/atom at 100 GPa. The enthalpy difference has about 50 meV/atom drop for $Ca_2UH_{18}$-*P*-3*m*1 at 140 GPa. The sudden change at 140 GPa in both $Ca_2UH_{18}$-*P*-3*m*1 and $CaU_3H_{32}$-*Pm*-3*m* is owing the convex hull modification caused by $CaH_6$ [Fig. 1(c)], illustrating that binary hydrides play critical role in the convex hull of Ca-U-H. Therefore, the synthesize of $CaUH_{12}$-*Cmmm* and $CaUH_{12}$-*Fd*-3*m* needs relatively severe conditions; $Ca_2UH_{18}$-*P*-3*m*1 is meta-stable and the phonon results suggest it could exist until 100GPa; $CaU_3H_{32}$-*Pm*-3*m* is also meta-stable and its dynamical stability at 90 GPa indicates the synthesizing potential below 1 mega-bar.

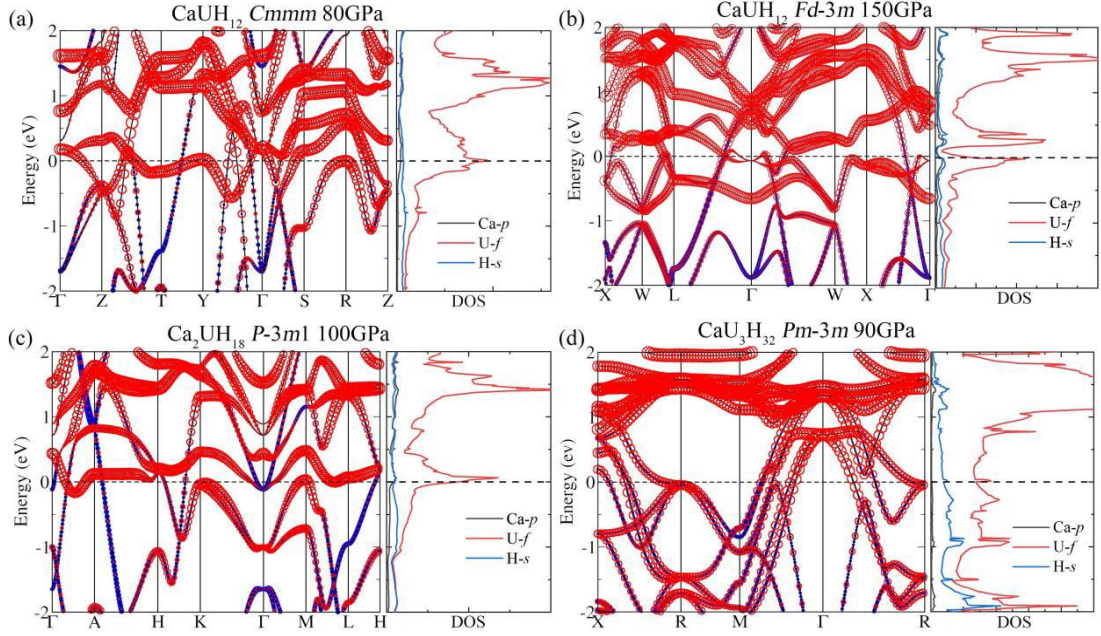

FIG. 3. The band structures and corresponding PDOS of the predicted structures under high pressure. The contribution of Ca, U and H are decorated in black, red and blue. The Fermi energy is set to zero.

Next, we calculated the electronic structures of the predicted structures. As depicted in Fig. 3, the valence bands and conduction bands overlap with each other around the Fermi energy in all the four band structures, illustrating typical metal characteristics. Among them, $CaUH_{12}$-$Cmmm$ has a near flat band along the Brillouin zone path $T$-$Y$; $CaUH_{12}$-$Fd$-$3m$ has Dirac-type alike crossing along the $L$-$\Gamma$ direction; $Ca_2UH_{18}$-$P$-$3m1$ has both flat band along $A$-$H$ direction and Dirac-type alike crossing in $\Gamma$-$A$ path; $CaU_3H_{32}$-$Pm$-$3m$ has a saddle point at $R$ point. These features correspond to the peaks in the partial density of states (PDOS), which may be beneficial for superconductivity. Besides, both the band structures and PDOS of the predicted structures indicate that the $f$ electrons of U atoms make the main contribution around the Fermi energy. This is distinct to the hydrogen dominant PDOS in Ca-H compounds, but analogous to the hydrogen-rich U-H compounds, such as $UH_8$ and $UH_9$ under high pressure [33]. Our calculation parameters for electronic structure followed Ref. 33 and Ref. 52. In addition, Ref. [56] checked different effective on-site Coulombic repulsion U in Dudarev et al. formulation [57], and pointed out that the influence of electronic correlation effect can be neglected in the hydrogen rich uranium hydrides.

To further explore the bonding condition in the predicted Ca-U-H structures under high pressure, we conducted COHP, integrated COHP (ICOHP) and electron localization function (ELF) calculations. According to the convention, the positive and negative values of -COHP stand for bonding and anti-bonding characteristics, and ICOHP can identify the covalent-like bonding. As shown in Fig. S4, U-H exhibits bonding characteristics below Fermi energy, while Ca-H displays anti-bonding fingerprint around -10 eV. The absolute ICOHP values of U-H is about one order greater than Ca-H, indicating that the U atoms have stronger bonding interactions with H atoms than Ca atoms. The ELF results are displayed in Fig. S5. There is almost no local charge between Ca and H, suggesting ionic bonding properties, while the ELF between U and H atoms indicates covalent bonding. The ELF results and COHP analysis indicate that the bonding properties of U-H in Ca-U-H resemble that in hydrogen rich uranium hydrides under high pressure [52].

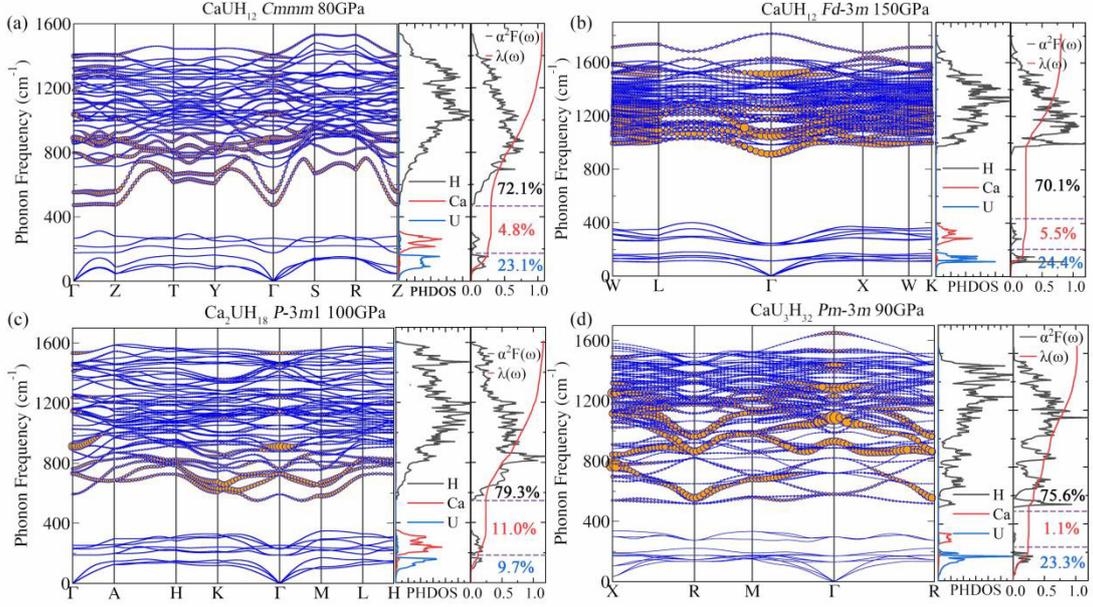

FIG. 4. The calculated phonon curves, PHDOS, Eliashberg spectral function $α^2F(ω)$, the electron-phonon integral $λ(ω)$ the predicted structures. The size of the orange solid dots represents the contribution to electron phonon coupling.

In Fig. 4, we calculated the phonon curves, projected phonon density of states (PHDOS), Eliashberg spectral function $α^2F(ω)$ and the electron-phonon integral $λ(ω)$ of the predicted structures under their lowest dynamically stable pressure. In the phonon spectrum and PHDOS, the vibration frequencies are divided into two parts. Ca and U atoms play crucial roles in the low-frequency part (0-400 $cm^{-1}$) and the H atoms contribute to the high-frequency region (>400 $cm^{-1}$). The main distribution of Eliashberg spectral function $α^2F(ω)$ is in the high-frequency region, in which the electron-phonon integral $λ(ω)$ has around 70-80% rising. The above results illustrate that Ca and U have moderate contribution to the total EPC constant and H atoms play dominant role in EPC. The EPC constants is 1.03, 0.71, 1.14 and 1.07 for $CaUH_{12}$-$Cmmm$ at 80 GPa, $CaUH_{12}$-$Fd$-$3m$ at 150 GPa, $Ca_2UH_{18}$-$P$-$3m1$ at 100 GPa and $CaU_3H_{32}$-$Pm$-$3m$ at 90 GPa respectively. The EPC constants of $CaUH_{12}$-$Fd$-$3m$ is relatively lower than that of other structures. Comparing with the phonon spectrum under 200 GPa [Fig. S1], we can observe several softened phonon modes in the 400-800 $cm^{-1}$ region of $CaUH_{12}$-$Cmmm$, $Ca_2UH_{18}$-$P$-$3m1$ and $CaU_3H_{32}$-$Pm$-$3m$ [Fig. 4(a), (c) and (d)]. We calculated the mode-resolved electron phonon coupling constants $λ_{q,ν}$ [58],

$$λ = \sum_{q,ν} λ_{q,ν} ω_{q,ν} \quad (1)$$

where $\lambda$ is the total electron-phonon coupling constant, $q$, $v$ are are the phonon vector and the index of the mode, respectively, and $\omega_{q,v}$ is the weight parameters. The size of the solid dots in Fig. 4 is the contribution of the vibration mode to electron-phonon coupling. These several softened phonon modes contribute 23.06% for $CaUH_{12}$-$Cmmm$ at 80 GPa, 22.26% for $Ca_2UH_{18}$-$P$-$3m1$ at 100 GPa and 22.16% for $CaU_3H_{32}$-$Pm$-$3m$ at 90 GPa in the integral $\lambda(\omega)$, respectively, indicating their high contribution to the total electron-phonon coupling. Since the U-derived DOS are comparable in these four predicted structures, the softened phonon frequencies are advantages to the enhancement of EPC constant.

Table II. The calculated superconducting properties of the predicted Ca-U-H structures. Two $T_c$ values were calculated with $\mu^*$ equal to 0.13 and 0.10, respectively.

| Phase | Space Group | Pressure (GPa) | $\lambda$ | $T_c$ (K) |
|---|---|---|---|---|
| $CaUH_{12}$ | $Cmmm$ | 80 | 1.03 | 50.6 |
|  |  |  |  | 58.6 |
| $CaUH_{12}$ | $Fd$-$3m$ | 150 | 0.71 | 24.3 |
|  |  |  |  | 31.4 |
| $Ca_2UH_{18}$ | $P$-$3m1$ | 100 | 1.14 | 57.5 |
|  |  |  |  | 65.8 |
| $CaU_3H_{32}$ | $Pm$-$3m$ | 90 | 1.07 | 43.5 |
|  |  |  |  | 50.1 |

To estimate the superconducting transition temperature $T_c$ of the predicted structures, we used Allen-Dynes modified McMillan formula.

$$T_c = \frac{\omega_{\log}}{1.2} \exp\left(\frac{-1.04(1+\lambda)}{\lambda - \mu^*(1+0.62\lambda)}\right) \quad (2)$$

where $\omega_{\log}$ is the logarithmic average frequency, and $\mu^*$ is the Coulomb pseudopotential for which we used the widely accepted values 0.10 and 0.13. The EPC constant $\lambda$ and $\omega_{\log}$ are defined as below.

$$\lambda = 2\int_0^\infty \frac{\alpha^2 F(\omega)}{\omega} d\omega \quad (3)$$

and

$$\omega_{\log} = \exp\left(\frac{2}{\lambda}\int \frac{d\omega}{\omega}\alpha^2 F(\omega)\ln(\omega)\right) \quad (4)$$

The calculated $T_c$ values for the predicted structures are listed in Table II. Among them, the $T_c$ values of the meta-stable phase $CaU_3H_{32}$-$Pm$-$3m$ reaches 50.1 K at 90 GPa, and $Ca_2UH_{18}$-$P$-$3m1$ reaches 65.8 K at 100 GPa, which approaches the temperature range of liquid nitrogen.

Furthermore, if we neglect the ternary compositions in the ternary convex hull, the enthalpy minimum is located at $CaH_4$ under 100 GPa, which implies that the thermodynamically stable ternary compositions may be along the triangle straight line from CaH$_4$ to UH. We expect to find more Ca-U-H ternary hydrides with lower thermodynamic pressure or higher $T_c$ in this energy surface region in the future work.

## Conclusion

In summary, we have explored the Ca-U-H ternary hydrides under high pressure using the machine learning graph theory accelerated structure searching combined with first-principles calculations. We predicted four ternary hydrides which all take the clathrate structures: CaUH$_{12}$-*Cmmm*, CaUH$_{12}$-*Fd-3m*, Ca$_2$UH$_{18}$-*P-3m*1 and CaU$_3$H$_{32}$-*Pm-3m*. CaUH$_{12}$-*Cmmm* and CaUH$_{12}$-*Fd-3m* may exist in annealing process after laser heating, Ca$_2$UH$_{18}$-*P-3m*1 and CaU$_3$H$_{32}$-*Pm-3m* are meta-stable and their $T_c$ values are above 40 K. Ca$_2$UH$_{18}$-*P-3m*1 and CaU$_3$H$_{32}$-*Pm-3m* have potential to be synthesized below 1 megabar. Besides, the *f* electrons in U atoms play crucial role in the electronic density of states and the softened vibration modes contributes significantly to the electron phonon coupling. Our work brings possibility of new superconducting ternary hydrides by introducing actinides into alkaline-earth metal hydrides. This could provide novel route in designing superconducting ternary hydrides and inspire future experimental studies.

## Acknowledgment

This work was supported by the National Natural Science Foundation of China (Grant Nos. 52272265, U1932217, 11974246, 12004252), the National Key R&D Program of China (Grant No. 2018YFA0704300), and the Shanghai Science and Technology Plan (Grant No. 21DZ2260400).